\documentclass[twocolumn,showpacs,preprintnumbers,amsmath,amssymb,aps,prl,nofootinbib]{revtex4}

\usepackage{graphicx}
\usepackage{dcolumn}
\usepackage{bm}
\usepackage{times}
\usepackage{multirow}
\usepackage{footmisc}
\usepackage{color}

\newcommand{\rhor}{\rho({\bf r})}
\newcommand{\Vext}{V_{\rm ext}({\bf r})}

\begin{document}

\title{Rhombic preordering on a square substrate}

\author{T. Neuhaus}
\author{M. Marechal}
\author{M. Schmiedeberg}
\author{H. L\"owen}
\affiliation{%
Institut f\"ur Theoretische Physik II: Weiche Materie,
Heinrich-Heine-Universit\"at D\"{u}sseldorf,
Universit{\"a}tsstra{\ss}e 1, D-40225 D\"{u}sseldorf, Germany}
\date{\today}

\begin{abstract}
A competition of incommensurate symmetries occurs whenever a system is forced to conform to an ordering that is different from the intrinsically preferred structure of the system itself. As a model system of such a competition, we study the rivalry between the triangular ordering of hard disks and the square symmetry induced by a periodic square substrate. By using density functional theory as well as Monte Carlo computer simulations, we determine the full phase behavior for the case of one particle per minimum. We observe a rhombic preordering structure preceding the hexagonal solid as a direct consequence of the competing symmetries. Furthermore, the square-rhombic transition is reentrant with increasing substrate interaction. Our predictions can be verified in experiments of colloids in laser fields.
\end{abstract}


\maketitle


If a monolayer of particles is adsorbed on a substrate with an incompatible symmetry, there is a competition between the phase that the adsorbed particle would form without substrate and the phase that reflects the symmetry of the substrate. As already described by Landau~\cite{Landau1937}, there must be a first order transition between phases with incompatible symmetries. Phase transitions induced by substrates are widely studied, for example on substrates with a one dimensional commensurate structure~\cite{LIF}, which enhances the triangular order that is present without substrate. On other substrates with more complex periodic or even aperiodic symmetry, new phases have been observed~\cite{Bechinger2001,Brunner2002,Reichhardt2002,Mangold2003,SQUARE,PFC,Schmiedeberg2008,AT,ElShawish2008}.
For small interactions between the particles and the substrate, the phase preferred by the particles in the absence of a substrate strength prevails, while the phase preferred by the substrate is enforced for
a strong substrate strength. In many systems, additional phases at intermediate substrate strength have been observed, for example for nanoparticles or micron-sized colloids \cite{Bechinger2001,Mangold2003,Schmiedeberg2008,AT}, for vortices in type-II superconductors~\cite{Reichhardt2001,Berdiyorov2009}, for adsorbed atoms (cf., e.g., \cite{Coppersmith1981,Grunze1983,SQUARE}) or molecules (for a recent review, see \cite{Arnold2012}). An intermediate phase also has been predicted from defect-mediated melting theory \cite{Nelson1979}. 
\begin{figure}[b!]
\center
\includegraphics[width=5.0cm]{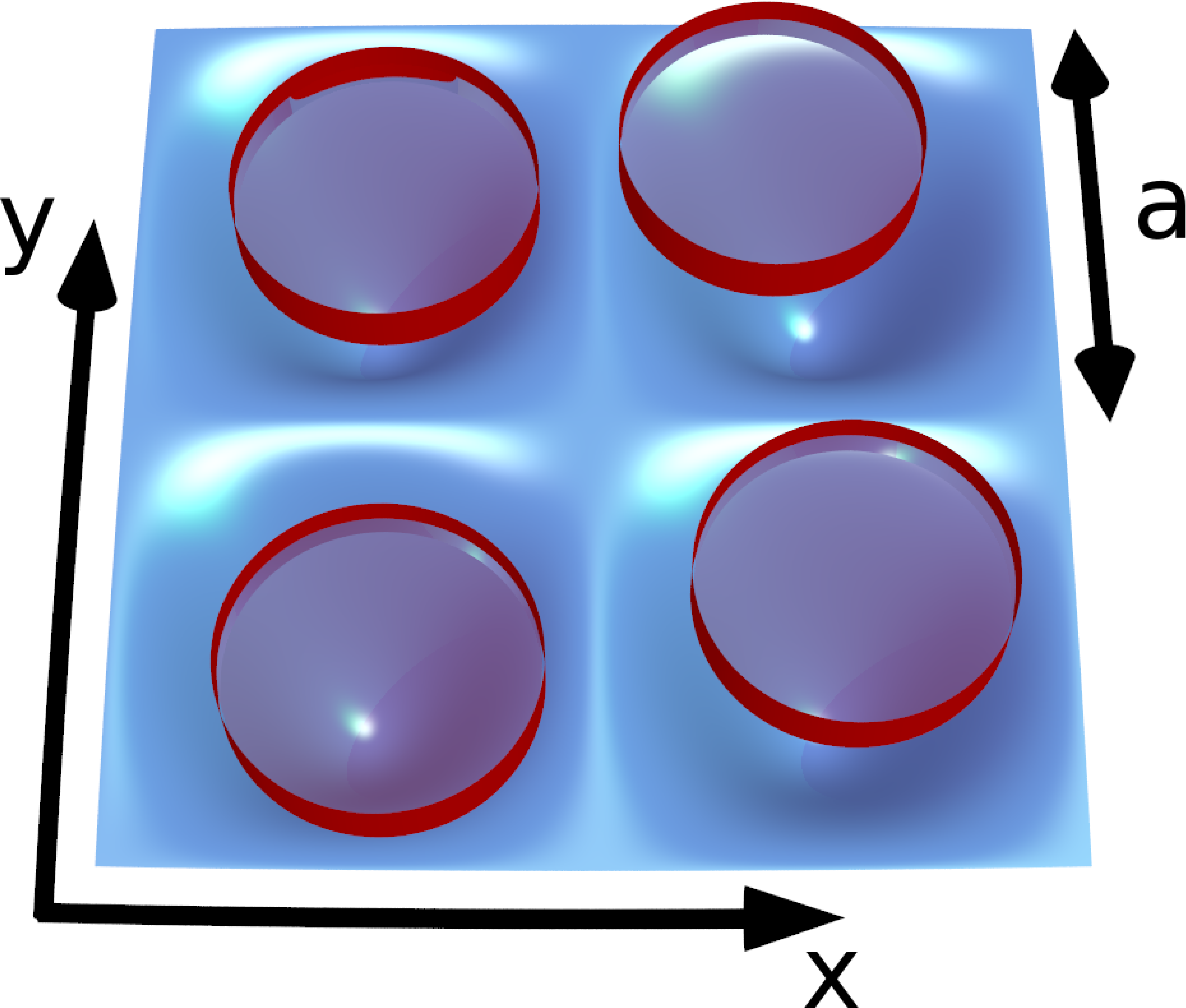}
\caption{\label{fig:scheme}
Schematic view of the square substrate potential of lattice constant $a$. The disks have diameter $\sigma$ and on average there is only one particle located at each minimum of the substrate.
}
\end{figure}

In this letter we study the competition of incompatible symmetries in a minimal model system. We consider hard disks adsorbed on a square substrate for a density that is chosen such that there is always one disk per minimum of the substrate. A schematic depiction of the model is shown in Fig.~\ref{fig:scheme}. Without the substrate potential, a system of hard disks at any temperature forms a triangular solid (with quasi-long range order)~\cite{Mitus1997} if the density is sufficiently high. Evidently, the disks prefer triangular order which is incompatible to the square symmetry of the substrate. By using a fundamental-measure density functional theory (DFT) as well as Monte Carlo simulations, we determine the phase behavior depending on the packing fraction and the strength of the particle-substrate interactions.
At low packing fraction (large wave length of the external potential), the system is in a modulated square fluid phase. When increasing the packing fraction, we observe a rhombic preordering as a consequence of
the incompatibility of the symmetries prior to the first order transition into the triangular phase at high packing fraction. In the rhombic ordering, the particles are still bound to the minima of the square substrate. However, every second row or every second column of particles is shifted in one direction, while the other rows or columns are displaced in the opposite direction. In this way, the rhombic structure allows for a larger mean distance between the particles acting as a precursor to the hexagonal solid, where the distance between nearest neighbors is maximal. Using the mean-field type DFT, we find a second-order square-fluid to rhombic transition, while with Monte Carlo simulations there is no first or second order transition but a continuous crossover from the modulated square symmetry to rhombic ordering.

Our results indicate a more general scenario in which two competing incommensurate structures lead to the emergence of a preordering structure which inherits properties from both competing phases. As a further result, a reentrant square-rhombic transition or crossover is found for increasing strength of the substrate interaction. 
Our model system can be realized experimentally by sterically-stabilized colloids on a flat surface or at an air-water interface exposed to a laser interference pattern with square symmetry, similar to previous experiments with colloidal particles where laser fields were used to model an external potential~\cite{LIF,Burns1990,Bechinger2001,Brunner2002,Mangold2003,AT}. The limit of strong square substrates has already been studied~\cite{Blaaderen1997,Bechinger2001,Reichhardt2002,ElShawish2008}, usually with the emphasis on the case of multiple colloids per minimum \cite{Bechinger2001,Reichhardt2002,Mangold2003,ElShawish2008}.

In our model, the hard disks have a diameter $\sigma$ which serves as length scale. The thermal energy $k_BT$ is the energy scale. The external potential is
\begin{equation}
 \Vext=V_0\bigg(1-\frac{1}{4}\bigg|\sum \limits_{j=1}^4e^{i{\bf k}_j\cdot{\bf r}}\bigg|^2\bigg),
 \label{eq:ext_pot}
\end{equation}
with the amplitude $V_0$, the four reciprocal lattice vectors $\{{\bf k}_j\}=\{(\pm 1,\pm1)\frac{\pi}{a}\}$, and the lattice constant $a$ of the substrate square lattice. We study the system at unit filling, i.e.\ there is one particle per square, see again Fig.~\ref{fig:scheme}.  Unit filling implies that the areal number density is $1/a^2$ which  immediately translates into the dimensionless area fraction $\eta= \pi \sigma^2/4a^2$ of the system.

In the following, we shall first present results for the phase diagram from fundamental-measure DFT~\cite{Rosenfeld1989,Tarazona2008,Roth2010,Roth2012} in the two-dimensional plane spanned by the area packing fraction $\eta$ and the substrate potential strength $V_0/k_BT$. Subsequently, we describe our Monte Carlo simulation results.

In DFT~\cite{Evans1979}, a grand canonical free energy functional $\Omega(T,\mu,A,[\rhor])$ is minimized with respect to the density profile $\rhor$ at fixed temperature $T$, chemical potential $\mu$ and given area of the system $A$. The functional $\Omega(T,\mu,A,[\rhor])$ is conveniently split into three parts,
\begin{equation}
 \Omega[\rhor]=\mathcal{F}_{\rm id}[\rhor]+\mathcal{F}_{\rm exc}[\rhor]+\mathcal{F}_{\rm ext}[\rhor],
 \label{eq:omega}
\end{equation}
where $\mathcal{F}_{\rm id}[\rhor]=k_BT\int {\rm d}{\bf r} \rhor \left[\ln \left(\Lambda^2 \rhor \right)-1\right]$ is the exactly known ideal gas functional which includes the (irrelevant) thermal wavelength $\Lambda$. $\mathcal{F}_{\rm exc}[\rhor]$ denotes the nontrivial excess free energy functional resulting from the interaction between the particles for which we adopt the recently developed fundamental-measure theory for hard disks~\cite{Roth2012}. The third term describes the coupling to the external potential of Eq.~\eqref{eq:ext_pot} and reads 
\begin{equation}
 \mathcal{F}_{\rm ext}[\rhor]=\int {\rm d}{\bf r} \rhor[\Vext-\mu].
\end{equation}
\begin{figure}
\center
\includegraphics[width=\columnwidth]{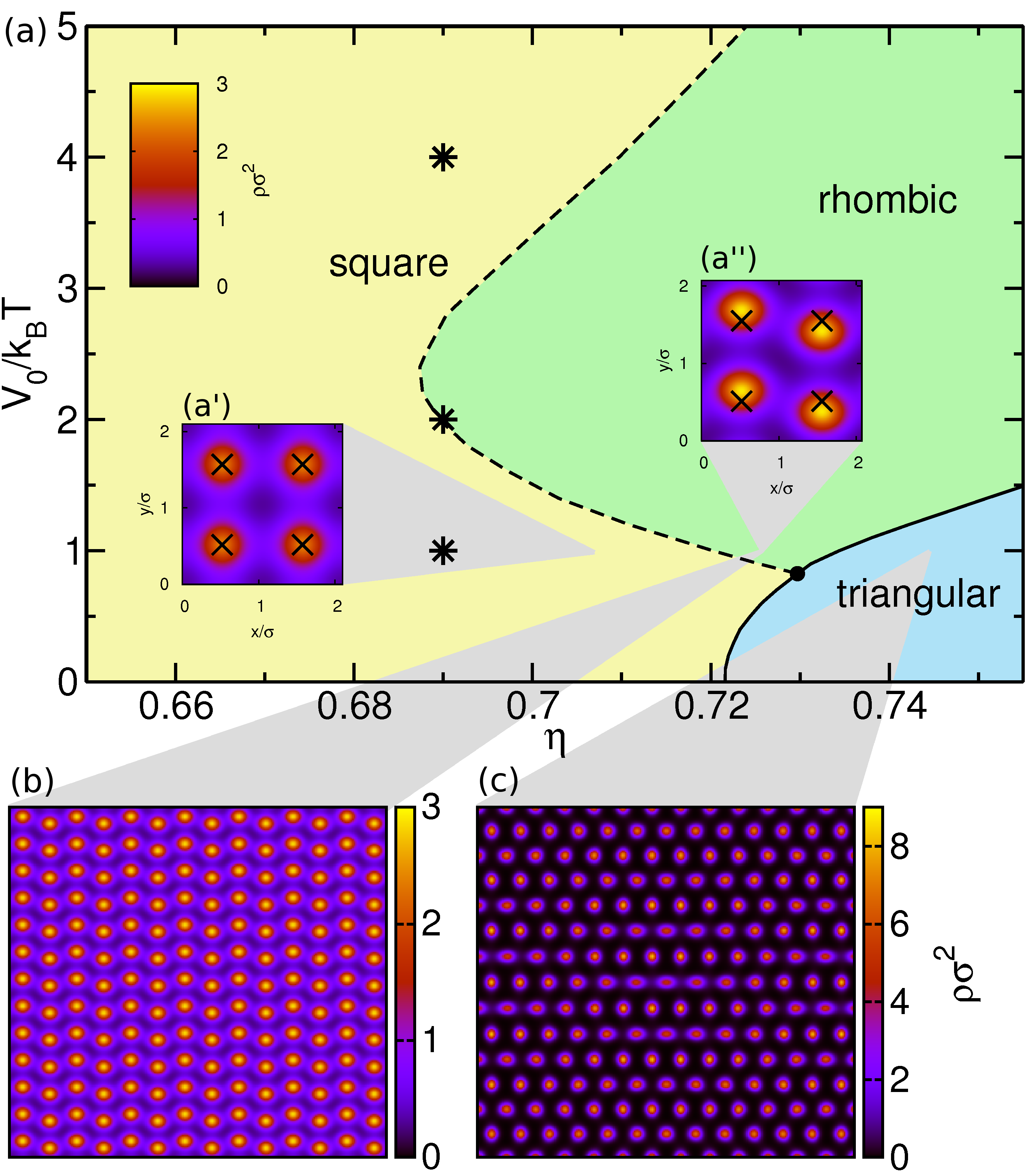}
\caption{\label{fig:phasediag}
(a) Phase diagram for hard disks on a square-substrate showing the three different phases, square modulated fluid, rhombic, and triangular crystal. The dashed line indicates the second order and the solid
line the first order phase transition, where the dot denotes the triple point. Density contour plots of the three phases are included obtained for a fixed external potential with $V_0/k_BT=1$ and packing fractions (a') $\eta=0.7069$, [(a''), (b)] $\eta=0.7257$ , and (c) $\eta=0.7383$. The stars indicate the simulation parameters for Fig.~\ref{fig:snapshots}.}
\end{figure}
We use a Picard iteration~\cite{Roth2010} scheme to minimize the grand-canonical free energy $\Omega$ freely on a fine grid. The chemical potential $\mu$ is used as a Lagrangian multiplier so that the average number of particles (i.e. the packing fraction $\eta$) in the system is fixed. We analyze the symmetry in the minimizing equilibrium density profile and obtain the phase diagram shown in Fig.~\ref{fig:phasediag}. For vanishing external potential $V_0=0$, we recover the bulk hard disk freezing transition from a disordered fluid to a hexagonal crystal. The freezing transition is first order in this approximation of the DFT~\cite{Roth2012}. Increasing the amplitude of the external potential $V_0$, particles in the fluid phase arrange on the square substrate and thus form a modulated fluid phase, see the density profile in Fig.~\ref{fig:phasediag}, where black crosses display the potential minima. Consequently, the phase transition shifts to higher packing fractions, as the external potential suppresses the triangular structure. For reduced amplitudes higher than $V_0 /k_BT=0.82$, a rhombic ordering of particles occurs. The rhombic structure can be thought of as a shift of even numbered rows of particles away from the lattice sites of the square lattice in one of the four possible directions, while odd numbered rows are shifted in the opposite direction. Thereby the square to rhombic transition can be in principle continuous, similar to a Martensitic transition \cite{Weiss1995}. In fact, our DFT studies show that the square-to-rhombic fluid transition is second order while the crystallization of either modulated fluid or rhombic phase into the triangular lattice is first order due to the incompatibility of the two structures. The resulting density contour plot of the triangular crystal (see Fig.~\ref{fig:phasediag}) shows a distorted hexagonal crystal.
The reentrant behavior can be understood by considering the interplay between entropy and the substrate potential: At low interactions with the substrate, the fluid is nearly unperturbed, except for a density modulation prescribed by $V_\text{ext}(\mathbf{r})$.  When the substrate interaction is not too strong, the entropy, which is mainly limited by interactions with the neighbors, can be increased by forming the rhombic structure, which has a larger mean inter-patricle distance. At even higher $V_0$, the fluctuations are suppressed to a point where the particles hardly interact with each other, such that they simply conform again with the structure enforced by the substrate.

\begin{figure}[t!]
\includegraphics[width=0.98\columnwidth]{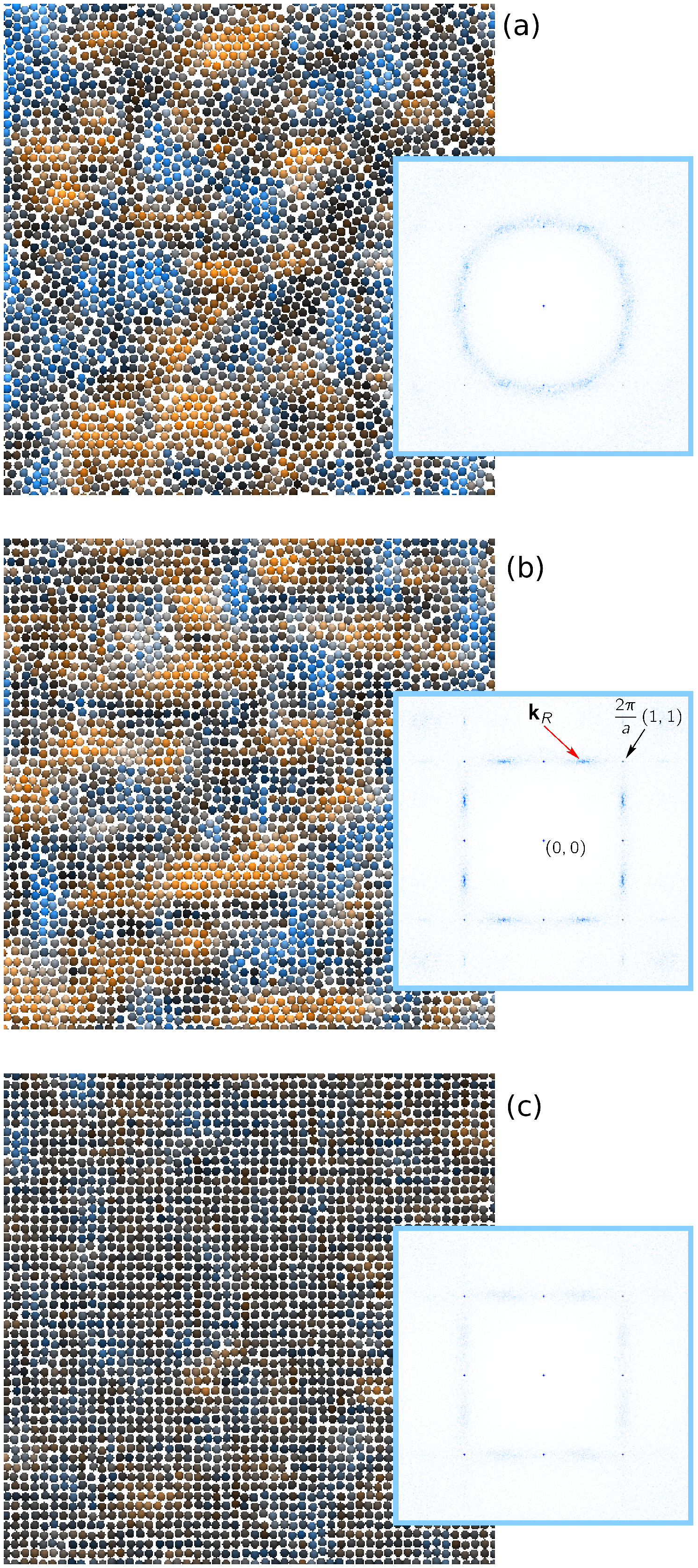}
\caption{ Typical configurations from Monte Carlo simulations for a system with $128^2$ particles (only a small section of the system is shown) and the corresponding scattering function (insets) for a packing fraction $\eta=0.69$ and external field strength (a) $V_0/k_B T=1$, (b) $2$ and (c) $4$ (as indicated in Fig.~\ref{fig:phasediag}). The particles are colored according to their local structure: black denotes no rhombic order, orange and blue denote perfect rhombic order with rows along the $x$ and $y$ directions, respectively. The scattering peak at $\mathbf{k}_R=(2\pi/a)\,(1,1/2)$ due to the rhombic pre-ordering is indicated in the inset of (b).
\label{fig:snapshots}}
\vspace*{-2em}
\end{figure}
In our Monte Carlo computer simulations, we fixed the number of particles $N$, area $A=N a^2$ and temperature $T$. Typical configurations for three $V_0$ values, along with the corresponding scattering function or structure factor $S(\mathbf{k})$~\cite{SI} are shown in Fig.~\ref{fig:snapshots} for a packing fraction $\eta=0.69$, well below the bulk phase transition. For a weak substrate potential, the system forms small clusters with hexagonal order, which are also found in the fluid for $V_0=0$. The field aligns the clusters, such that rows of particles occupy rows of minima in the substrate potential in the $x$ and $y$ directions. The resulting scattering profile has twelve-fold symmetry as predicted by Nelson~\cite{Nelson1979}, which is the result of a superposition of two scattering profiles each exhibiting hexagonal local order, where the second profile is rotated by 90${}^\circ$ compared to the first. For $V_0/k_BT=2$, the structure is completely different: the competition between the substrate potential, that is minimized by a square structure, and the free volume that is larger for triangular local structure, results in a locally rhombic structure. Formation of \emph{equilibrium} 
rhombic clusters with their rows along the two different directions causes the rhombic order to be finite ranged. The shifted rows cause large peaks at $\mathbf{k}_R=(2\pi/a)\,(1,1/2)$ in between the main peaks. Finally for large $V_0/k_BT=4$, the particles are forced to be near one of the minima of the substrate potential, which precludes rhombic ordering. Comparable results from DFT can be found in the Supplemental Material~\cite{SI}.

\begin{figure}[b!]
\includegraphics[width=\columnwidth]{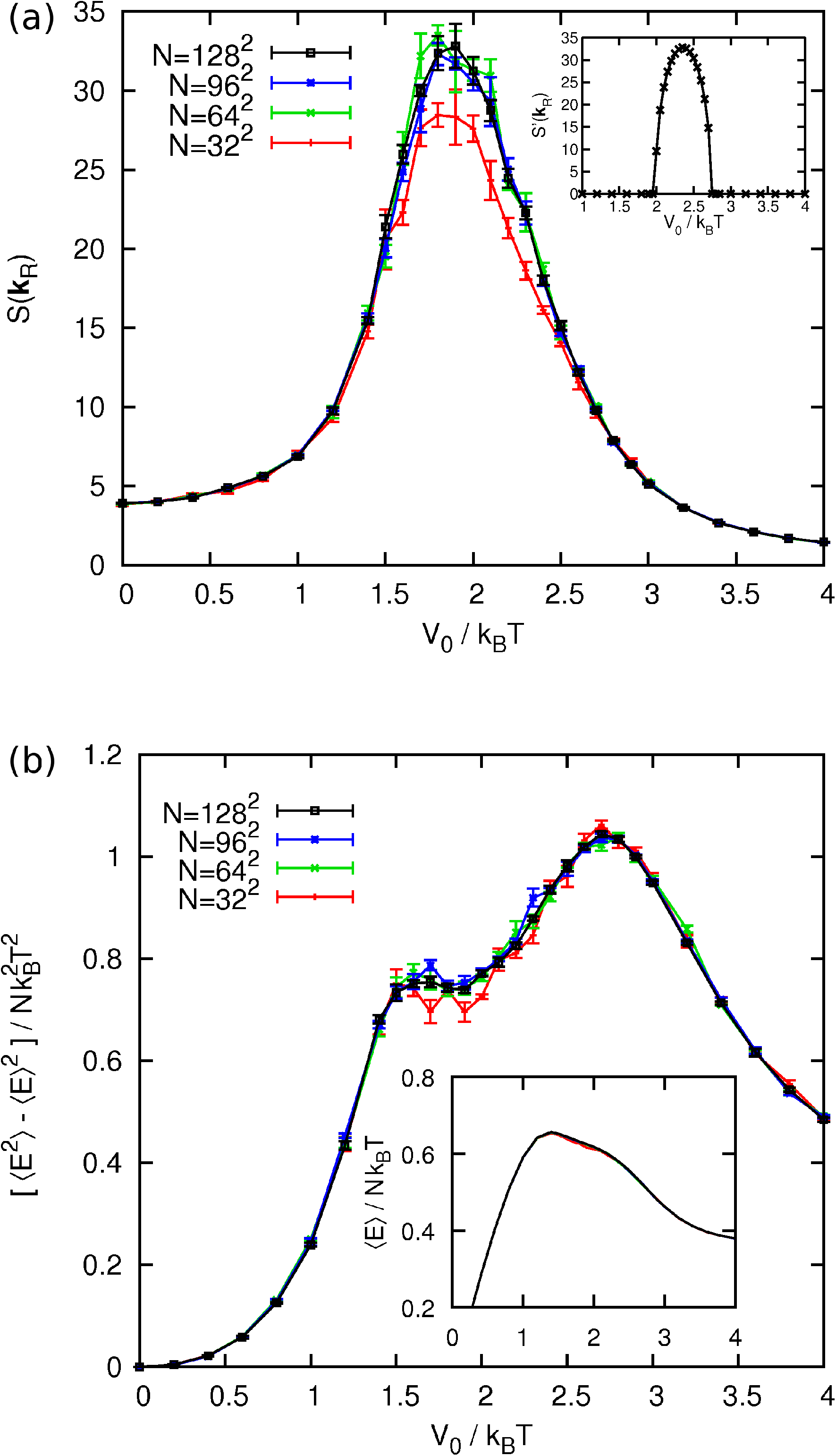}
\caption{Monte Carlo simulation results for packing fraction $\eta=0.69$ and for system sizes $N=32^2$, $64^2$, $96^2$ and $128^2$. (a) The scattering function $S(\mathbf{k})$ at the $\mathbf{k}$ vector that is indicative of rhombic order, $\mathbf{k}_R\equiv 2\pi/a (1,1/2)$, as a function of the amplitude $V_0$ of the external field. $S'({\bf k}_R)$ obtained from DFT in the inset. (b) The fluctuations of the energy divided by the system size $N$ as a function of $V_0$. The inset shows the average energy per particle, for which the error is of order of the line thickness.
\label{fig:sim_res}}
\end{figure}

It is clear from the scattering profile that the peak height at $\mathbf{k}_R$ is a good order parameter for rhombic ordering. We show $S(\mathbf{k}_R)$ as a function of $V_0$ for different system sizes in Fig.~\ref{fig:sim_res}(a). For every system size considerably larger than the largest rhombic cluster size in the system (\emph{i.e.} $N>32^2$), the scattering function has no finite size dependence, an unambiguous sign of exponential decay of the rhombic positional order with distance. In the inset of Fig.~\ref{fig:sim_res}(a), we show the corresponding peak height $S'(\mathbf{k}_R)$ as resulting from a Fourier analysis of the density field in DFT; for details see~\cite{SI}. It is nonzero in the rhombic phase, because the rhombic positional order is long-ranged in DFT. We see that the rhombic order, measured by $S(\mathbf{k}_R)$, in the simulations is much larger around the area where the DFT predicts a stable rhombic phase, so we would expect to find phase transitions where the $S(\mathbf{k}_R)$ changes most sharply in the simulations. In Fig.~\ref{fig:sim_res}(b), we show the average potential energy, which would show a discontinuity at a first order phase transition, and the mean squared energy fluctuations divided by the number of particles $N$, which would show a peak that increases with the system size at a second order phase transition. Neither quantities show an indication of a phase transition, although the variance of the energy shows system size-independent peaks around the $V_0$ values, where the rhombic order sharply changes. Clearly, we have found a very unusual transition, which has a clear structural signature, but no thermodynamic footprint. Furthermore, as mentioned before, the Monte Carlo simulations confirm the reentrant behavior found in the DFT. 

In conclusion, we have employed DFT calculations to study the phase behavior of hard disks on a square substrate and found rhombic preordering which was confirmed in our Monte Carlo simulations. Therefore, we have shown that DFT can be successfully applied to freezing on incommensurate structures. Only the order of the phase transition between the square to rhombic transition was underestimated by the mean-field like DFT.

Our results show how freezing occurs on a substrate with a structure that is incompatible with the symmetry of the bulk crystal. We found a new intermediate ordering that is a compromise between the two
symmetries. The model that we studied applies to the competition between square and hexagonal structures where the new emerging intermediate phase is rhombic, but we anticipate that this is a general scenario. Hence, we expect a comparable phase behavior for other competing symmetries, as shown for a rectangular substrate in the Supplemental Material~\cite{SI}. However, it is not immediately self-evident how a possible preordering, which corresponds to the rhombic phase on the square lattice, might look like on substrates with other symmetries. Therefore, an interesting question arising from our results is whether the intermediate phases reported for substrates with incompatible symmetries, e.g., the phases with 20-fold bond orientational order \cite{Schmiedeberg2008} or the Archimedean-like tiling phases \cite{AT} on substrates with decagonal quasicrystalline symmetry, are reminiscent of a preordering before a first order transition between completely incompatible symmetries. Furthermore, it will be interesting to explore other model systems where the incompatibility of different symmetries is induced by the boundaries~\cite{Neser1997,Ricci2007,Wilms2012} or is due to an incommensurate filling fraction of the substrate minima (see, e.g. \cite{Bechinger2001,Mangold2003}), and to study the modification of friction due to the different phases~\cite{Vanossi2012}.

\begin{acknowledgments}
We thank M. Oettel for helpful discussions.
This work was supported by the DFG via SPP 1296 and SFB TR6.
\end{acknowledgments}

\end{document}


\title{\Large Rhombic preordering on a square substrate:\\ Supplemental Material}
\author{T. Neuhaus}
\author{M. Marechal}
\author{M. Schmiedeberg}
\author{H. L\"owen}

\affiliation{Institut f\"ur Theoretische Physik II, Heinrich-Heine-Universit\"at D\"usseldorf,
  Universit\"atsstra{\ss}e 1, 40225 D\"usseldorf, Germany }

\maketitle

\vspace*{-1em}

\section{Simulation details}

The scattering function $S(\mathbf{k})$ (or structure factor) shown in the insets of Fig.~\RefFigSnapshots\ of the main text is calculated using a fast Fourier transform  $\hat{\rho}_\text{inst}(\mathbf{k})$ of the \emph{instantaneous} density profile of the corresponding snapshot binned on a two-dimensional grid: $S_\text{inst}(\mathbf{k})=\lvert \hat{\rho}_\text{inst}(\mathbf{k}) \rvert^2/N$. The method of plotting the scattering profile results in a slight smoothing of the peaks. As a result, the peaks at $\mathbf{k}=(2\pi/a)\,(\pm 1,\pm 1)$, which are actually essentially point-like, obtain a finite width rendering them visible. For the results in Fig.~\RefFigSimRes(a) of the main text, a more precise calculation of the average scattering function was used:
\begin{equation}
S(\mathbf{k}_R)=\frac{1}{N} \left\langle \left\lvert \sum_{n=1}^N e^{\mathrm{i} \mathbf{k}_R \cdot \mathbf{r}_n} \right\rvert^2 \right\rangle,
\label{eq:S_kR}
\end{equation}
which was averaged over multiple configurations during the simulation and over the different symmetry related $\mathbf{k}_R$ vectors. The coloring of the particles in Fig.~\RefFigSnapshots\ of the main text is performed as follows, the degree to which particle $i$ are orange and light blue is equal to $S_x^{(R)}(i)$ and $S_y^{(R)}(i)$ repectively, where 
\begin{equation}
S_{x}^{(R)}(i)\equiv \max_{\sigma=\pm 1}\frac{1}{N_i} \sum_j \cos\left(\tfrac{2\pi}{a} (\hat{x} + \sigma \hat{y}/2 ) \cdot [\mathbf{r}_j-\mathbf{r}_i]\right)
\end{equation}
and $S_{y}^{(R)}(i)$ is given by the same expression with $x$ and $y$ interchanged. The sum over $j$ is limited to the $N_i$ particles that are in a square of width $5 a$ centered around the position $\mathbf{r}_i$ of particle $i$.

\section{Direct comparison of the results from simulations and theory}

\begin{figure}[tb]
\includegraphics[width=0.9\textwidth]{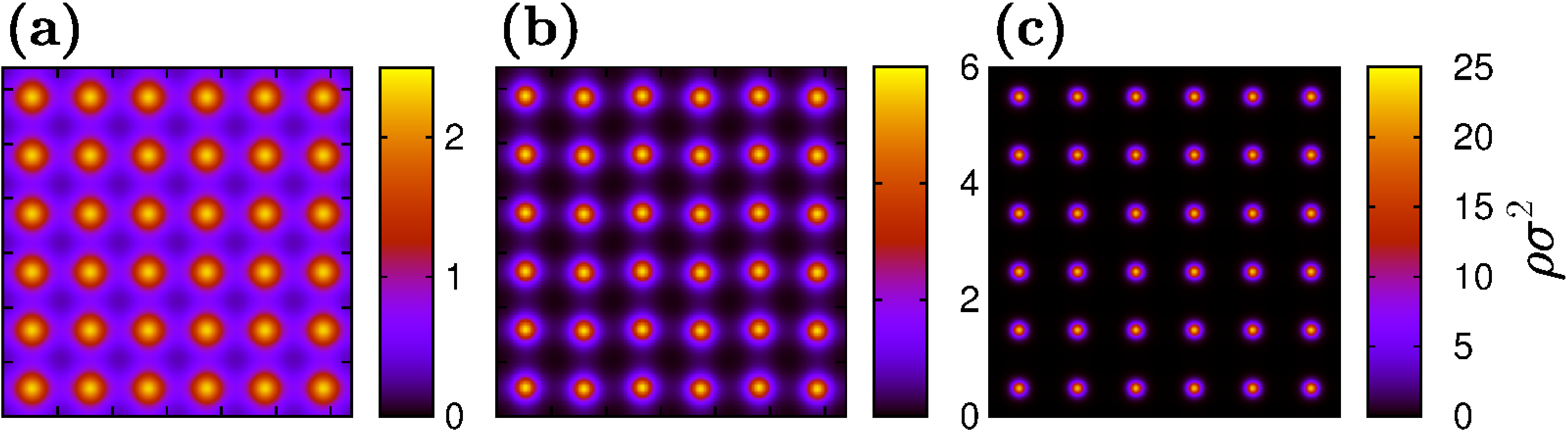}
\caption{Density contour plots of 6x6 squares for fixed packing fraction $\eta=0.69$ and external field strength (a) $V_0/k_BT=1$, (b) $V_0/k_BT=2$ , and (c) $V_0/k_BT=4$, similar to those indicated by stars in Fig.~2.}
\label{fig:supmat_1}
\end{figure}

In order to clarify the similarities of simulations and theory, first the density profiles obtained from DFT are computed for the state points marked by black stars in Fig.~2(a) and the results are shown in Fig.~\ref{fig:supmat_1}. Similar to the simulation snapshots shown in Fig.~3, particles in the square phase at high interaction strength $V_0$ are strongly pinned or in other words, the density peaks are very high (see Fig.~\ref{fig:supmat_1}(c)). At low $V_0$ (Fig.~\ref{fig:supmat_1}(a)), the square phase is a modulated fluid with much lower density peaks.
In DFT, the rhombic phase spans the whole system (see Fig.~\ref{fig:supmat_1}(b)), whereas a simulation snapshot consists of regions of rhombic and regions of square symmetry. As the state point for the rhombic phase is very close to the phase boundary, the rhombic structure is hardly visible but it can be confirmed by analyzing the profiles.

In analogy to the analysis of the results from our MC simulations, a structure factor related quantity can be calculated. The Fourier transform of the mean density distribution $\hat{\rho}({\bf k})$ is used as
\begin{equation}
 S'({\bf k})=\frac{\lvert\hat{\rho}({\bf k})\rvert^2}{S_0},
\end{equation}
conatining a normalization factor $1/S_0$. Here, the average and the Fourier transform are taken in opposite order compared with Eq.~\ref{eq:S_kR}. The results are in good agreement with those presented in Fig.~3. There is an intermediate peak at ${\bf k}_R=\tfrac{2\pi}{a}(1/2,1)$ only in the rhombic phase. In the simulations, the shift of particles from the square to rhombic phase occurs in the $x$ as well as the $y$ direction forming different clusters. In contrast, DFT yields only one way of dislocation. Thus, the intermediate peaks can only occur in one of the two possible directions. In the inset of Fig.~4(a), we plot the peak height at ${\bf k}_R$ as a function of $V_0/k_BT$. All $S'({\bf k})$ are normalized with a factor $1/S_0$ such that the peak height obtained from DFT agrees with that of the black curve in Fig.~4(a) resulting from our simulations.
\begin{figure}[t]
\includegraphics[width=0.9\textwidth]{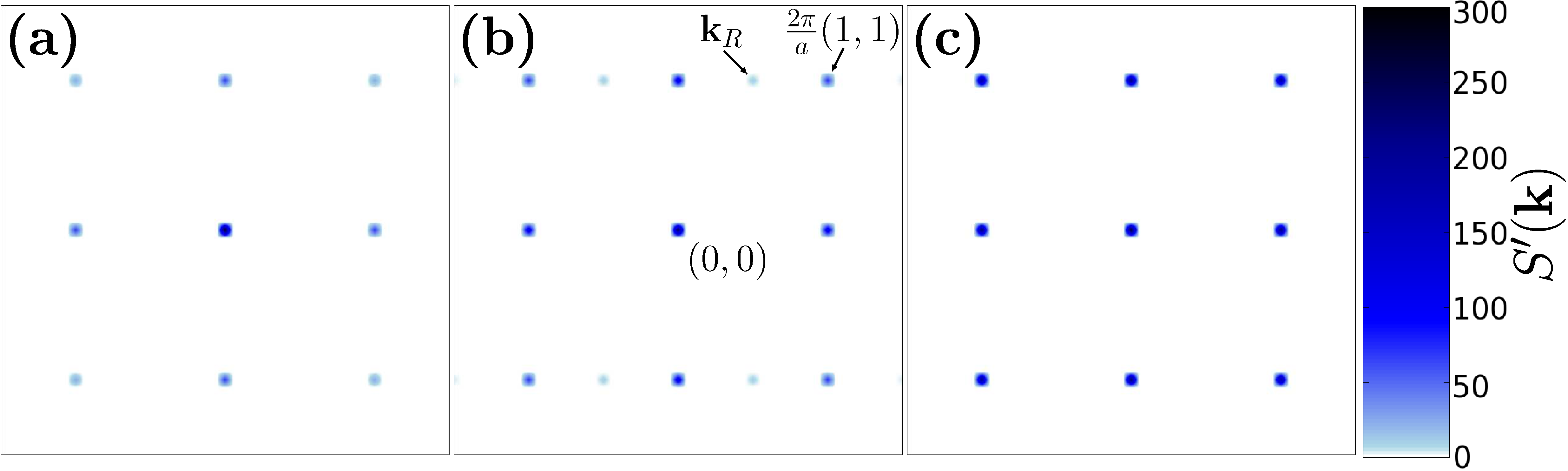}
\caption{Structure factor $S({\bf k})$ obtained from DFT at the state points indicated in Fig.~\ref{fig:supmat_1}}
\label{fig:supmat_2}
\end{figure}

\begin{figure}[htb]
\includegraphics[width=0.5\textwidth]{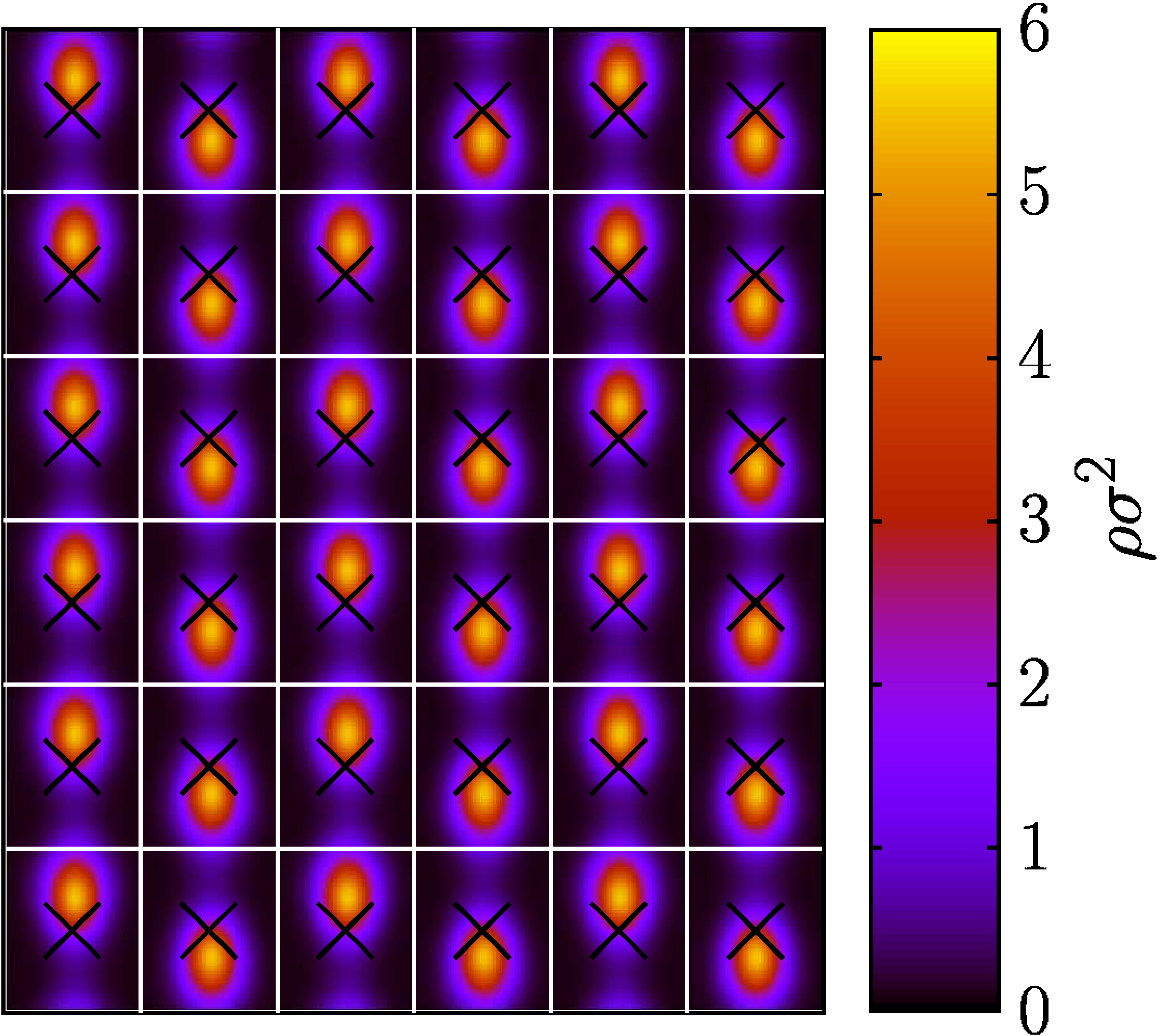}
\caption{Rhombic phase at packing fraction $\eta=0.69$ on a rectangular pattern with size ratio 1:1.2 and external field strength $V_0/k_BT=1.5$ indicated by black crosses.}
\label{fig:supmat_3}
\end{figure}

Additional to the analysis of a square substrate, an external potential with rectangular symmetry can be induced for a system with on average one particle per substrate minimum. This setup also yields a rhombic preordering as shown in a density contour plot in Fig~\ref{fig:supmat_3}.
